\documentclass[11pt,a4paper]{article}
\usepackage{cite}
\usepackage{amssymb}
\usepackage{amsmath}
\usepackage{graphicx}
\usepackage{relate}
\newtheorem{lemma}{Lemma}
\newtheorem{theorem}{Theorem}
\newtheorem{corollary}{Corollary}

\newcommand{\qed}{\hfill\ensuremath{\Box}\medskip\\\noindent}

\newcommand{\fullversion}[1]{#1}
\newcommand{\conference}[1]{}

\newcommand{\ceil}[1]{\left\lceil{#1}\right\rceil}

\newcommand\unfinished[1]{}

\title{
Fast evaluation of union-intersection expressions}

\author{Philip Bille\thanks{Computational Logic and Algorithms Group, IT University of Copenhagen, Denmark.} \and Anna Pagh$^*$ \and Rasmus Pagh$^*$}
\date{}

\begin{document}
\maketitle

\begin{abstract}
We show how to represent sets in a linear space data structure such 
that expressions involving unions and intersections of sets
can be computed in a worst-case efficient way. This problem has applications in e.g.~information retrieval and database systems. We mainly consider the RAM model of computation, and sets of machine words, but also state our results in the I/O model. On a RAM with word size $w$, a special case of our result is that the intersection of $m$ (preprocessed) sets, containing $n$ elements in total, can be computed in expected time $O(n (\log w)^2 / w + km)$, where $k$ is the number of elements in the intersection. If the first of the two terms dominates, this is a factor $w^{1-o(1)}$ faster than the standard solution of merging sorted lists. We show a cell probe lower bound of time $\Omega(n/(w m \log m)+ (1-\tfrac{\log k}{w}) k)$, meaning that our upper bound is nearly optimal for small~$m$.
Our algorithm uses a novel combination of approximate set representations and word-level parallelism.
\end{abstract}


\section{Introduction} 
Algorithms and data structures for sets play an important role in computer science. For example, the
relational data model, which has been the dominant database paradigm for decades, is based on set representation and manipulation. Set operations also arise naturally in connection with database queries that can be expressed as a boolean combination of simpler queries. For example, search engines report documents
that are present in the intersection of several sets of documents, each corresponding to a word in the
query. If we fix the set of documents to be searched, it is possible to spend time on preprocessing all sets, to decrease the time for answering queries.

The search engine application has been the main motivation in several recent works on computing set intersections~\cite{conf/soda/DemaineLM00,conf/soda/BarbayK02,conf/icalp/ChiniforooshanFM05}. All these papers assume that elements are taken from an ordered set, and are accessed through comparisons. In particular, creating the canonical representation, a sorted list, is the best possible preprocessing in this context. The comparison-based model rules out some algorithms that are very efficient, both in theory and practice. For example, if the preprocessing produces a hashing-based dictionary for each set, the intersection of two sets $S_1$ and $S_2$ can be computed in expected time $O(\min(|S_1|,|S_2|))$. This is a factor $\Theta(\log(1+\max(\frac{|S_1|}{|S_2|},\frac{|S_2|}{|S_1|})))$ faster than the best possible worst-case performance of comparison-based algorithms.

In this paper we investigate non-comparison-based techniques for evaluating expressions involving unions and intersections of sets on a RAM. (In the search engine application this corresponds to expressions using {\sc AND} and {\sc OR} operators.) Specifically, we consider the situation in which each set is required to be represented in a linear space data structure, and propose the {\em multi-resolution set representation}, which is suitable for
efficient set operations. We show that it is possible in many cases to achieve running time that is sub-linear in the total size of the input sets and intermediate results of the expression. For example, we can compute the intersection of a number of sets in a time bound that is sub-linear in the total size of the sets, plus time proportional to the total number of input elements in the intersection.
In contrast, all previous 
algorithms that we are aware of take at least linear time in the worst case over all possible input sets, even if the output is the empty set. The time complexity of our algorithm improves as the word size $w$ of the RAM grows. While the typical word size of a modern CPU is 64 bits, modern CPU design is {\em superscalar\/} meaning that several independent instructions can be executed in parallel. This means that in most cases (with the notable exception of multiplication) it is possible to {\em simulate\/} operations on larger word sizes with the same (or nearly the same) speed as operations on single words. We expect that word-level parallelism may gain in importance, as a way of making use of the increasing parallelism of modern processor architectures.

\subsection{Related work} 

\subsubsection{Set union and intersection}

The problem of computing intersections and unions (as well as differences) of sorted sets was recently
considered in a number of papers (e.g.~\cite{conf/soda/DemaineLM00,conf/soda/BarbayK02}) in an {\em adaptive\/} setting.
A good adaptive algorithm uses a number of comparisons that is close (or as close as possible) to
the size of the smallest set
of comparisons that determine the result. In the case of two sorted sets, this is the number of interleavings when merging the sets. In the worst case this number is linear in the size of the sets, in which
case the adaptive algorithm performs no better than standard merging. However, adaptive algorithms are able to exploit ``easy'' cases to achieve smaller running time. Mirzazadeh in his thesis~\cite{Mirzazadeh} extended this line of work
to arbitrary expressions with unions and intersections. These results are incomparable to those obtained in this paper: Our algorithm is faster for most problem instances, but the adaptive algorithms are faster in certain cases. It is instructive to consider the case of computing the intersection of two sets of size $n$ where the size of the intersection is relatively small.
In this case, an optimal adaptive algorithm is faster than our algorithm only if the number of {\em interleavings\/} of the sorted lists (i.e., the number of sublists needed to form the sorted list of the union of the sets) is less than around $n/w$.

Another idea that has been studied is, roughly speaking, to exploit asymmetry. Hwang and Lin~\cite{SICOMP::HwangL1972} show that merging two sorted lists $S_1$ and $S_2$ requires $\Theta(|S_1| \log(1+\frac{|S_2|}{|S_1|}))$ comparisons, for $|S_1|<|S_2|$, in the worst case over all input lists. 
This is significantly less than $O(|S_1|+|S_2|)$ if $|S_1|\ll |S_2|$. This result was generalized to 
computation of general expressions involving unions and intersections of sets by Chiniforooshan et 
al.~\cite{conf/icalp/ChiniforooshanFM05}. Given an expression, and the sizes of the input sets, their algorithm uses a number
of comparisons that is asymptotically equal to the minimum number of comparisons required in the
{\em worst case\/} over all such sets.\footnote{After personal communication with the authors, we have had confirmed that the algorithm described in~\cite{conf/icalp/ChiniforooshanFM05} is not optimal in certain cases. Specifically, it does not always compute the union of sets in the optimal bound. However, the authors have informed us that the algorithm can be slightly modified to remove this problem.}
The bounds stated in~\cite{conf/icalp/ChiniforooshanFM05} do not involve the size of the 
output, meaning that they pessimistically assume the output to be the largest possible, given the
expression and the set sizes. In contrast, our bounds will be {\em output sensitive}, i.e., involve also
the size of the result of the expression. 
We further compare our result to that of~\cite{conf/icalp/ChiniforooshanFM05} in section~\ref{sec:results}.

\subsubsection{Approximate set representations}

There has been extensive previous work on approximate set representations, mainly motivated by
applications in networking and distributed systems~\cite{BM:02}. Much of this work builds upon the
seminal paper on Bloom filters~\cite{Bloom:1970:STT}. A Bloom filter for a set $S$ is an approximate
representation of $S$ in the sense that for any $x\not\in S$ the filter can be used to determine that
 $x\not\in S$ with probability close to $1$. However, for an $\epsilon$ fraction of elements not in $S$,
called {\em false positives}, the Bloom filter is consistent with a set that includes these elements.
The advantage of allowing some false positives, rather than storing $S$ exactly, is that the space usage
drops to around $O(n\log(1/\epsilon))$ bits, practically independent of the size of the universe of which $S$
is a subset. Two Bloom filters for sets $S_1$ and $S_2$ can be combined to form a Bloom filter for
$S_1\cap S_2$ (resp.~$S_1\cup S_2$), in a very simple way: By taking bitwise AND (resp.~OR) of the
data structures.

Bloom filters have been used in connection with computation of relational joins, which are 
essentially multiset intersections, in the I/O model of computation. 
The idea is to use a Bloom filter for the smaller set to efficiently
find most elements of the larger set that are {\em not\/} in the intersection. If the Bloom filter can fit
into internal memory, this is a highly efficient procedure for reducing the amount of data that needs
to be considered in the join. The algorithm presented in this paper also uses approximate set
representations to eliminate elements that will not contribute to the result. However, using Bloom
filters does not appear to yield an efficient solution, essentially because the information pertaining to
a particular element of $S$ is distributed across the data structure. This makes it hard to locate the
set of input elements represented by a particular Bloom filter. Instead, we use the approximate
set representation of Carter et al.~\cite{MR80h:68037} (see also~\cite{bloom}), which consists of storing, in a compact way, the image of the set under a universal hash function.

\subsection{Setup and results}\label{sec:results}

We consider fully parenthesized expressions with binary operators. That is, we have a rooted binary tree with input sets at the leaves and internal nodes corresponding to union and intersection operations. Given the sizes of all input sets, we may associate with any node $v$ two numbers (notation from~\cite{conf/icalp/ChiniforooshanFM05}): 
\begin{itemize}
\item $\psi(v)$ is the maximum possible number of elements in the subexpression rooted at $v$. (Can be computed bottom-up by summing child values at union nodes, and choosing the minimum child value at intersection nodes.)
\item $\psi^*(v)$ is the maximum possible number of elements in the subexpression rooted at $v$ that can appear in the final result. This is the minimum value of $\psi(v)$ on the path from $v$ to the root.
\end{itemize}
We denote by $V$ the set of nodes in the expression (internal as well as leaves), and let $v_0$ denote the root node.

\begin{theorem}\label{thm:main}
Given suitably preprocessed sets of total size $n$, we can compute the value of an expression with binary union and intersection operators in expected time $O(k' + \sum_{v\in V} \lceil\tfrac{\psi^*(v)}{w} \log^2(\tfrac{n w}{\psi^*(v)}) \rceil)$, where $k'$ is the number of occurrences in the input of elements in the result. Preprocessing of a set of size $n_1$ uses linear space and expected time $O(n_1 \log w)$.
\end{theorem}

Theorem~\ref{thm:main} requires some effort to interpret.
We will first state some special cases of the result, and then discuss the general result towards the end of the section. 
It is not hard to see that the terms in the sum of Theorem~\ref{thm:main} corresponding to intersection nodes do not affect the asymptotic value. That is, we could alternatively sum over the set of leaf nodes and union nodes in the expression. In the case where the expression is an intersection of $m$ sets we can further improve our algorithm and analysis to get the following result:

\begin{theorem}\label{thm:intersect}
Given $m$ preprocessed sets of total size $n$, we can compute the intersection of the sets in expected time $O(n \log^2 w/w+ km)$, where $k$ is the number of elements in the intersection.
\end{theorem}

We show the following lower bound, implying that the time complexity of Theorem~\ref{thm:intersect} is within a factor $(\log w)^2 m \log m$ of optimal, assuming $w=(1+\Omega(1))\log n$. Our lower bound applies to the class of functions whose union-intersection expression has an intersection operation on any root-to-leaf path (an element needs to be in at least two input sets to appear in the result). Note that if there is a path consisting of only union operations, there exists a set where all elements must be included in the result, so this requirement is no serious restriction.

\begin{theorem}\label{thm:lower}
Let $f$ be a function of $m$ sets given by a union-intersection expression with an intersection node on any root-to-leaf path.
For integers $n$ and $k\leq n/m$, any (randomized) algorithm in the cell probe model that takes representations of sets $S_1,\dots,S_m \subseteq \{0,1\}^w$,
where $\sum_i |S_i| \leq n$ and $|f(S_1,\dots, S_m)| \leq k$, and computes $|f(S_1,\dots,S_m)|$ 
must use expected time at least $\Omega(n/(w m \log m)+ (1-\tfrac{\log k}{w}) k)$ on a worst-case input. The lower bound holds regardless of how the sets are represented.
\end{theorem}

\fullversion{
It is possible to (coarsely) bound the sum of Theorem~\ref{thm:main} in terms of the total size of the input sets and the $\cup$-depth of the expression (maximum number of unions on a root-to-leaf path):

\begin{corollary}
Given $m$ preprocessed sets of total size $n$, we can compute the value of an expression of $\cup$-depth $d$ with binary union and intersection operators in expected time $O(m+k'+\tfrac{n}{w} (d+ \log w)^2)$, where $k'$ is the number of occurrences in the input of elements in the result.
\end{corollary}
}

Possibly the best way of understanding the general
result in Theorem~\ref{thm:main} is to compare the complexity to the comparison-based algorithm of~\cite{conf/icalp/ChiniforooshanFM05}. Though it might not result in the best running time for our algorithm, we make the comparison in
the case where any group of adjacent union operators is arranged as a perfectly balanced tree in the expression tree (we could modify our algorithm to always make this change to the expression). The algorithm of~\cite{conf/icalp/ChiniforooshanFM05} takes an expression where operators have unbounded degree, and where union and intersection nodes alternate. It can be applied in our setting by combining groups of adjacent union and intersection operators. The time usage is {\em at least\/} $\Omega(k' + \sum_{v\in V} {\psi^*(v)})$ (in fact, the complexity also involves a logarithmic factor on each term, but it is not easily comparable to the factor in our result).
Thus, if the word length is sufficiently large, e.g.~$w=(\log n)^{\omega(1)}$, our algorithm gains a factor $w^{1-o(1)}$ compared to~\cite{conf/icalp/ChiniforooshanFM05}.

We observe that all our results immediately imply nontrivial results in the I/O model~\cite{MR90k:68029}.
For the upper bounds, this is because any RAM algorithm can be simulated in the same I/O bound as long as $w$ is bounded by the number of bits in a disk block. In other words, if $B$ is the number of words in a disk block, we can get I/O bounds by replacing $w$ by $Bw$ in the results. In fact, the power of $2$ in the bounds can be reduced to $1$ in this setting, as the I/O model does not count the cost of computation. Our lower bound also holds in the I/O model, with $w$ replaced by $Bw$, independently of the size of internal memory. (The same proof applies.)

\subsection{Technical overview.}

Our results are obtained through non-trivial combination of several known techniques.
We use the idea of Carter el al.~\cite{MR80h:68037} to obtain an approximate representation of a set
by storing a set $h(S)$ of hash function values rather than the set $S$ itself. Storing the approximation
in a na{\"i}ve way (using at least $\log n$ bits per element) does not lead to a significant speedup
in general. Instead, a compact representation of the set $h(S)$ is needed.
We use a bucketed set representation, as in the dictionary of Brodnik and Munro~\cite{BM2}, to get a compact
representation of $h(S)$ that is suitable for word-parallel set operations.
Specifically, we show how set operations on small integers packed in words can be
efficiently implemented, using ideas from~\cite{AH1997, AHNR1995}. This allows us to quickly approximate the intersection of any two sets in the sense that we get a compressed list of references to the elements in the intersection plus a small fraction of the elements not in the intersection. To compute the intersection we compute the intersection of the subsets of ``candidates'' in the standard way, using hashing. The generalization to the case of expressions involving arbitrary unions and intersections is an extension of this idea, using a variant of a technique from~\cite{conf/icalp/ChiniforooshanFM05} to keep the sizes of the sets we have to deal with as small as possible. Our lower bound is shown by a reduction to multi-party communication complexity.


\section{Main algorithm and data structure}\label{sec:mainalgorithm}

In this section we present most of our algorithm and data structure, postponing the material on word-level parallelism to Section~\ref{sec:succinctsets} (which is used as a blackbox in this section). 
Specifically, we
show how to reduce the problem of performing unions and intersections on sets of words to the problem of performing these operations on sets from a smaller universe. 
\fullversion{Due to space constraints, the time and space analysis is placed in Appendix~\ref{app:analysis}.}%
\conference{Due to space constraints we refer to the technical report version of this paper~\cite{arXiv-intersect} for the time and space analysis.}

\subsection{Overview of special case: Intersection}

We first present the main ideas in the case where the expression is an intersection of $m$ sets.
The basis of the approach is to map elements of $\{0,1\}^w$ to a smaller universe using a hash 
function $h$, and compute the intersection $H=h(S_1)\cap\dots\cap h(S_m)$. Now, if $x\in S_1\cap \dots \cap S_m$ then $h(x)\in H$.
On the other hand, if $x\not\in S_1\cap \dots \cap S_m$ then, if $h$ is suitably chosen, we will
have $h(x)\not\in H$ with probability close to 1. Thus, we can regard $H$ as representing a good approximation of $S_1\cap\dots\cap S_m$. In particular, if we compute the sets 
$S'_i = \{ x\in S_i \; | \; h(x)\in H\}$, $i=1,\dots,m$, we expect that $S'_i$ does not contain many elements
of $S_i \backslash (S_1\cap\dots\cap S_m)$. Since $S_i \supseteq S'_i \supseteq S_1\cap\dots\cap S_m$ we can compute the intersection of $S_1,\dots,S_m$ as $S'_1\cap\dots\cap S'_m$ --- using a standard linear time hashing-based algorithm. The challenge of this approach is to keep the cost of
computing $H$ and the sets $S'_i$ low. We store preprocessed, compressed representations of
the sets $h(S_i)$ using only $O(\log w)$ bits per hash value, which allows us to compute $H$ in time that is sub-linear in the size of the input sets. The elements of $S'_i$ are extracted in additional time $O(|S_i|)$. The details of these steps appear in sections~\ref{sect:multiresolution} and~\ref{sec:succinctsets}.
Readers mainly interested in the case of computing a single intersection may skip the description of the
general case in the next subsection.

\subsection{The general case}

In the rest of the paper we let $f$ denote the function of $m$ input sets given by the expression to be evaluated. Since $f(S_1,\dots,S_m)$ is monotone in the sense that adding an element to an input set can never remove an element from $f(S_1,\dots,S_m)$ we have that for any $x\in f(S_1,\dots,S_m)$ it holds that $h(x)\in f(h(S_1),\dots,h(S_m))$. This means we can compute $f(S_1,\dots,S_m)$ by the following steps:
\begin{enumerate}
\item Compute $H=f(h(S_1),\dots,h(S_m))$.
\item For all $i$ compute the set $S'_i = \{ x\in S_i \; | \; h(x)\in H\}$.
\item Compute $f(S'_1,\dots,S'_m)$ to get the result.
\end{enumerate}
We will show how, starting with a suitable, compressed representation of the sets 
$h(S_1),\dots,h(S_m)$, we
can efficiently
perform the first two steps such that the sets $S'_i$
are significantly smaller than the $S_i$ in the following sense: {\em Most\/} of the elements that do not 
occur in $f(S_1,\dots,S_m)$ have been removed. This means that, except for negligible terms, the time for performing the third
step, using the standard linear time hashing-based algorithm, depends on the number of input elements in the output rather than on the size of the input.
Conceptually, the first step computes the expression on approximate representations of the sets
$S_1,\dots,S_m$. Then the information extracted from this is used to create a smaller problem instance
with the same result, which is then used to produce the answer. 

Assume for now that $h$ is given, and that we have access to data structures for $h(S_1),\dots,h(S_m)$.
The details on how to choose $h$ appear in Section~\ref{sect:multiresolution}.
The computation of $f(h(S_1),\dots,h(S_m))$ is done bottom-up in the expression tree in the same order as the
algorithm of~\cite{conf/icalp/ChiniforooshanFM05}: For an intersection node $v$ we first recursively process the child subtree whose root has the smallest value of $\psi^*$ --- the children of union nodes are processed recursively in arbitrary order. We adopt another idea of~\cite{conf/icalp/ChiniforooshanFM05}: If the set computed for the subtree rooted at $v$ has size more than $2\psi^*(v)$, we reduce the size of the set to at most $\psi^*(v)$ by computing the intersection with the smallest child set of an intersection node on the
path from $v$ to the root. Observe that this will only remove elements that are not in the output. Due to
the way we traverse the expression tree, the relevant child set will already have been computed. For every node $v$ in the expression tree, we store the result $\mathcal{I}_v$ of the subexpression rooted at $v$.

For the root node $v_0$ define $\mathcal{I}'_{v_0} = \mathcal{I}_{v_0}$.
To compute the sets $S'_i$ we first traverse the tree top-down and compute for every non-root node $v$ the intersection $\mathcal{I}'_v = \mathcal{I}'_v \cap \mathcal{I}'_{p(v)}$, where $p(v)$ is the parent node of $v$. Observe that, by induction, $\mathcal{I}'_v = \mathcal{I}_v\cap f(h(S_1),\dots,h(S_m))$.
We will see that
the time for this procedure is dominated by the time for computing $f(h(S_1),\dots,h(S_m))$. At the end we have computed $h(S'_i)=f(h(S_1),\dots,h(S_m)) \cap h(S_i)$ for all $i$. All that remains is to find the corresponding elements of $S'_i$, which is easily done by looking up the hash function values in a
hash table that stores $h(S_i)$ with the corresponding elements of $S_i$ as satellite information.

Finally, we compute $f(S'_1,\dots,S'_m)$ by first identifying all duplicate elements in the sets (by inserting them in a common hash table), keeping track of which set each element comes from. Then for each element decide whether it is in the output by evaluating the expression. This can be done in time proportional to the number of occurrences of the element: First annotate each leaf and intersection node in the expression tree with the nearest ancestor that is an intersection node. Then compute the set corresponding to each intersection node bottom-up. The time spent on an intersection node is bounded by the total size of the sets at intersection nodes immediately below it, but the intersection of these sets has size at most half of the total size. This implies the claimed time bound by a simple accounting argument.

\subsection{Data structure}\label{sect:multiresolution}

The best choice of $h$ depends on the particular expression and size of input sets. For example, when
computing the intersection $S_1 \cap S_2$ we want the range of $h$ to have size significantly 
larger than the smaller set ($S_1$, say). This will imply that most elements in $h(S_2\backslash S_1)$ 
will not be in $h(S_1)$, and there will be a significant reduction of the problem instance in step~2 of the main algorithm. 
On the other hand, the
time and space usage grows with the size of the range of the hash function used, so it should be chosen
no larger than necessary. In conclusion, to be able to choose the most suitable one in a given situation, we wish to store the image of every set under several hash functions, differing in the size of their range.
The images of the set under various hash functions can be thought of as representations of the set at
different resolutions. Hence, we name our data structure the {\em multi-resolution set representation}.
\fullversion{As we show in Appendix~\ref{app:analysis}, it suffices to use a hash function with range $\{0,1\}^r$, where }%
\conference{As we show in the full version of this paper~\cite{arXiv-intersect}, it suffices to use a hash function with range $\{0,1\}^r$, where}
$r=\log n + O(\log w)$ and
$n$ is the total size of the input sets.

The hash functions will all be derived from a single ``mother'' hash function $h^*$, a strongly universal hash function~\cite{CW,fast_universal} with values in the range $\{0,1\}^{w}$. This is a global hash function that is shared for all sets.
The hash function $h_r$, for $1\leq r\leq w$ is defined by 
$h_r(x)=h^*(x)\text{ div } 2^{w-r}$, where ``div'' denotes integer division (we use the natural correspondence between bit strings and nonnegative integers). 
Note that $h_r$ has function values of $r$ bits. To store $h_r(S)$ for a particular set $S$,
$r\geq \log |S| + 1$,
requires $|h_r(S)| (r-\log |h_r(S)| + \Theta(1))$ bits, by information theoretical arguments. Since we
may have $|h_r(S)|=|S|$ the space usage could be as high as $|S| (r-\log |S| + \Theta(1))$. Note that
the required space per element is constant when $r\leq \log |S| + O(1)$, and then grows linearly with $r$.

If we store $h_r(S)$ for all $r$, $\log |S| < r \leq w$, the space usage may be $\Omega(w)$ times 
that of storing $S$ itself. To achieve linear space usage we store $h_r(S)$ only for selected values 
of $r$, depending on $|S|$, namely 
$r\in\{ \lceil \log |S|\rceil + 2^i \; | \; i=0,1,2,\dots,\lfloor \log(w-\log |S|)\rfloor \}$. 
These sets are stored using the bucketed set representation of Section~\ref{sec:succinctsets} which gives a space
usage for $h_r(S)$ of $O(|S|(r-\log |S|+\log w))$ bits.
To get the representation of $h_r(S)$ for arbitrary $r$ we access the stored representation of $h_{r'}$, 
where $r'>r$, and throw away the $r'-r$ least significant bits of its elements (see Section~\ref{sec:succinctsets} for details). Choosing $r'$ as small as possible minimizes the time for this step.
We build the bucketed set representation of the largest value of $r$ in $O(|S|)$ time by hashing, and
then apply Lemma~\ref{lem:bucketconvert} iteratively to get the structures for the lower values of $r$.

The final thing we need is a hash table that allows us to look up a value $h_r(x)$ and retrieve the
element(s) in $S$ that have this value of $h_r$. This can be done by using the $\lceil \log |S|\rceil$
most significant bits of $h_r$ as index into a chained hash table. Since the values of these bits are common for all $h_r$, $\log |S| < r \leq w$, we only need to store a single hash table. Note that the
size of the hash table is $\Theta(|S|)$, which means that the expected lookup time is constant.




\section{Bucketed and packed sets}\label{sec:succinctsets}

We describe two representations of sets of elements from a small universe and provide efficient algorithms for computing union and intersection in the representations. Proofs of the lemmas in this
\fullversion{section can be found in Appendix~\ref{app:lemmas}.}%
\conference{section can be found in the full version~\cite{arXiv-intersect}.}

\subsection{Packed sets}\label{sec:packedsets}

Given a parameter $f$ we partition words into $k = w/(f+1)$ substrings, called \emph{fields}, numbered from right to left. The most significant bit of a field is called the \emph{test bit} and the remaining $f$-bits are called the \emph{entry}. A word is viewed as an array $A$ capable of holding up to $k$ bit strings of length $f$. If the $i$th test bit is $1$ we consider the $i$th field to be \emph{vacant}. Otherwise the field is \emph{occupied} and the bit string in the $i$th entry is interpreted as the binary encoding of a non-negative integer. If $|A| > k$ we can represent it in $\ceil{|A|/k}$ words; each storing up to $k$ elements. We call an array represented in this way a \emph{packed array with parameter $f$} (or simply \emph{packed array} if $f$ is understood from the context). For our purposes we will always assume that fields are capable of storing the total number of fields in a word, that is, $f \geq \log k$. In the following we present a number of useful ways to manipulate packed arrays.

Suppose $A$ is a packed array containing $x$ occupied fields. Then, \emph{compacting} $A$ means moving all the occupied fields into the first $x$ fields of $A$ while maintaining the order among them.
\begin{lemma}[Andersson et al.~\cite{AHNR1995}]\label{lem:packedcompact}
A packed array $A$ with parameter $f$ can be compacted in $O\left(|A|\ceil{f^2/w}\right)$ time.
\end{lemma}

Let $X = x_1, \ldots, x_m$ be a sequence of $f$-bit integers. If $X$ is given as a packed array with parameter $f$, such that the $i$th field, $1\leq i \leq m$, holds $x_i$, we say that $X$ is a \emph{packed sequence with parameter $f$}. We use the following result:
\begin{lemma}[Albers and Hagerup~\cite{AH1997}]\label{lem:packedmerge}
Two sorted packed sequences $X_1$ and $X_2$ with parameter $f$ can be merged into a single sorted packed sequence in $O\left((|X_1|+|X_2|)\ceil{f^2/w}\right)$ time.
\end{lemma}

We refer to a sorted, packed sequence of integers as a \emph{packed set}.
\begin{lemma}\label{lem:packedoperations}
Given packed sets $S_1$ and $S_2$ with parameter $f$, the packed sets $S_1 \cup S_2$ and $S_1 \cap S_2$ with parameter $f$ can be computed in $O\left((|S_1| + |S_2|)\ceil{f^2/w}\right)$ time.
\end{lemma}

\subsection{Bucketed sets}
Let $S$ be a set of $l$-bit integers. For a given parameter $b \leq l$ we partition $S$ into $2^b$ subsets, $S_0, \ldots, S_{2^b - 1}$, called \emph{buckets}. Bucket $S_i$ contain all values in the range $[2^{i(l-b)}, 2^{(i+1)(l-b)} - 1]$, and therefore all values in $S_i$ agree on the $b$ most significant bits. Hence, to represent $S_i$ it suffices to know the $b$ most significant bits together with the set of the $l-b$ least significants bits. We can therefore compactly represent $S$ by an array of length $2^b$, where the $i$th entry points to the packed set (with parameter $l-b$) of the $l-b$ least significant bits of $S_i$. We say that $S$ is a \emph{bucketed set with parameter $b$} if it is given in this representation. Note that such an encoding of $S$ uses $O(2^bw +  |S|(l-b))$ bits. As above, we assume that fields in packed sets are capable of holding the number of fields in a word, that is, we assume that $(l-b) \geq \log w - \log(l-b+1))$ in any bucketed set. 
We need the following results to manipulate bucketed sets.

\begin{lemma}\label{lem:bucketconvert}
Let $S$ be a bucketed set of $l$-bit integers with parameter $b$. Then,
\begin{enumerate}
  \item Given an integer $b'$ we can convert $S$ into a bucketed set with parameter $b'$ in time
  $O\left(2^{\max(b,b')} + |S|\ceil{(l - \min(b,b'))^2/w}\right)$.
  \item Given an integer $b < x \leq l$ we can compute the bucketed set $S' = \{j \:\mathrm{div}\: 2^{x} \mid j \in S\}$ of $l-x$ bit integers with parameter $b$ in $O\left(2^b + |S|\ceil{(l - b)^2/w}\right)$ time.
\end{enumerate}
\end{lemma}

Let $S$ be a bucketed set of $l$-bit integers with parameter $b$. We say that $S$ is a \emph{balanced bucketed set} if  $b$ is the largest integer such that $b \leq \log |S| - \log w$. Intuitively, this choice of $b$ balances the space for the array of buckets  and the packed sets representing the buckets. Since $l \geq \log |S|$ the condition implies that
$l - b \geq l - \log |S| + \log w \geq \log w - \log (l-b+1)$. Hence, the field length of the packed sets representing the buckets in $S$ is as required. Also, note that the space for a balanced bucketed set $S$ is $O(2^bw + |S|(l-b)) = O(|S|(l- \log |S| + \log w))$. 

\begin{lemma}\label{lem:bucketoperations}
Let $S_1$ and $S_2$ be balanced bucketed sets of $l$-bit integers. The balanced bucketed sets $S_1 \cup S_2$ and $S_1 \cap S_2$ can be computed in time $$O\left((|S_1| +|S_2|)\ceil{(l - \log(|S_1| + |S_2|) + \log w)^2/w}\right) \enspace .$$
\end{lemma}
If $l = \Theta(\log (|S_1| + |S_2|))$ Lemma~\ref{lem:bucketoperations} provides a speedup by a factor of $w/ \log^2 w$.


\unfinished{
\subsection{Observations XXX}

Number of bits read is factor log w less than implied by time. Could speed up algorithm by factor
log w by special instructions.

Could base algorithm on real sizes of intermediate results (rather than worst-case estimates) by
sampling (in a consistent way using hashing) all sets, and evaluate the function on the samples.

}

 
\section{Lower bound}\label{sec:lower}

In this section we show Theorem~\ref{thm:lower}. The proof uses known bounds from {\em $t$-party communication complexity}, where $t$ communicating players are required to compute a function of $n$-bit strings $x_1,\dots,x_t$, where $x_i$ is held by player $i$, using as little communication as possible. We consider the {\em blackboard model\/} where a bit communicated by one player is seen by all other players, and consider the following binary functions:
\begin{description}
\item[EQ($x_1,x_2$)] which has value 1 iff $x_1=x_2$. (Here $t=2$.)
\item[DISJ$_{n,t}(x_1,\dots,x_t$)] which has value 1 iff there is no position where two bit strings $x_i$ and $x_j$ both have a 1 (i.e., all pairs are ``disjoint''). We consider this problem under the {\em unique intersection assumption}, where either all pairs are disjoint, or there exists a single position where all bit strings have a 1. We allow the protocol to behave in any way if this is not the case.
\end{description}
Solving {\bf EQ} exactly requires communication of $\Omega(n)$ bits, for both deterministic and randomized protocols~\cite{STOC::Yao1979,KusNis:cc}. That is, the trivial protocol where one player communicates her entire bit string is optimal. Chakrabarti at al.~\cite{conf/coco/ChakrabartiKS03} showed that solving {\bf DISJ$_{n,t}$} exactly requires $\Omega(n/(t\log t))$ bits of communication in expectation, even under the unique intersection assumption and when the protocol is randomized.

Our main observation is that if sets $S_1,\dots,S_t$ have been independently preprocessed, 
we can view any algorithm that computes $f(S_1,\dots,S_t)$ as a communication protocol where each player holds a set. Whenever the algorithm accesses the representation of $S_i$
it corresponds to $w$ bits being sent by player $i$. Formally, given any (possibly randomized) algorithm that computes $|f(S_1,\dots,S_t)|$, where $S_1,\dots,S_t$ have been individually preprocessed in an arbitrary way, we derive
communication protocols for {\bf EQ} and {\bf DISJ$_{n,t}$}, and use the lower bounds for these problems to conclude a lower bound on the expected number of steps used by the algorithm. We note that this
reduction from communication complexity is different from the reduction from asymmetric communication
complexity commonly used to show data structures lower bounds.

Let $n$ and $k$, $1\leq k\leq n/t$, denote integers such that the algorithm correctly computes $|f(S_1,\dots,S_t)|$ provided that the sum of sizes of the sets is at most $n+1$, and that $|f(S_1,\dots, S_t)| \leq k$. Let $\tau$ denote the number of cell probes on a worst-case input of this form.
Given vectors $x_1,\dots,x_t \in\{0,1\}^n$ satisfying the unique intersection assumption, we consider the sets $S_i=\{ j \;|\; x_i \text{ has a 1 in position $j$}\}$ and their associated representations (which could be chosen in a randomized fashion). Observe that the total size of the sets is at most $n+1$, and that $|f(S_1,\dots,S_t)|=0$ if and only $S_1,\dots,S_t$ are disjoint (using the assumptions on $f$). By simulating the algorithm on these representations, we get a communication protocol for {\bf DISJ} using $\tau w$ bits in expectation. By the lower bound on {\bf DISJ$_{n,t}$} we thus have $\tau w=\Omega(n/(t\log t))$ on a worst case input, i.e., $\tau=\Omega(n/(w t\log t))$ cell probes are needed.

Consider the function $f'(S_1,S_2)=f(S_1,\dots,S_1,S_2)$. Clearly, a lower bound on the cost of computing $f'$ applies to $f$ as well.
We denote by $\binom{\{0,1\}^w}{k}$ the set of subsets of $\{0,1\}^w$ having size $k$.
Let $q=\lfloor\log_2 |\binom{\{0,1\}^w}{k}|\rfloor$, and let $\phi$ be any injective function from $\{0,1\}^q$ to $\binom{\{0,1\}^w}{k}$. Given two vectors $x,y\in\{0,1\}^q$ we consider the sets $S_1=\phi(x)$ and $S_2=\phi(y)$, which satisfy $|f'(S_1,S_2)| \leq k$ and $(t-1)|S_1|+|S_2|\leq n$. 
Since $\phi$ is injective, we have that
$x=y$ iff $|f'(S_1,S_2)|=k$. Thus, similar to above we get a communication protocol for {\bf EQ} that uses $\tau w$ bits in expectation on a worst-case input. By the lower bound on {\bf EQ} we have $\tau =\Omega(q/w)$, implying that $\tau=\Omega(k (w-\log_2 k) / w)$. The maximum of our two lower bounds
is a factor of at most two from the sum stated in the theorem, finishing the proof.


\section{Conclusion and open problems}

We have shown how to use two algorithmic techniques, approximate set representations and
word-level parallelism, to accelerate algorithms for basic set operations. Potentially, the results (or techniques) could have a number of applications in problem domains such as databases (relational, textual,\dots) where some preprocessing time (indexing) may be invested to keep the cost of 
queries~low.

It is an interesting problem whether our results can be extended to handle non-monotone set operators such as set difference. The technical problem here is that one would have to deal with two-sided errors in the estimates of the intermediate results. 
 
 \medskip
 
{\bf Acknowledgement.} We thank Mikkel Thorup for providing us useful insight on the use of word-level parallelism on modern processors.


\small
\bibliographystyle{abbrv}
\bibliography{intersect,my}

\fullversion{
\newpage

\appendix

\section{Analysis of our algorithm}\label{app:analysis}

\subsection*{Running time}

We now sketch the proof of Theorem~\ref{thm:main}, using results on the set representation described in 
Section~\ref{sec:succinctsets}. Observe that the worst case for our algorithm occurs when any
intermediate result of a node $v$ has size $\psi(v)$, so we need only consider this case. 
The intersection operations performed
when the subresult of a node $v$ has size at least $2\psi^*(v)$ can be regarded as ``free'', since at least half of the elements will be removed. We can thus charge the cost of this against the cost of computing the removed elements.
It follows from Lemma~\ref{lem:bucketoperations} that the time for the remaining parts of the
first two steps in the main algorithm use time $O(\sum_{v\in V} \lceil\tfrac{\psi^*(v)}{w} \log^2(\tfrac{\psi(v_0) w}{\psi^*(v)}) \rceil)$. 

The main part of the analysis is to bound the number of elements in
$S'_1,\dots, S'_m$ that are not part of the final result. We will show that in expectation there are
$O(n/w)$ such elements, implying that the time spent on these elements in the third step of the main algorithm is negligible. This will finish the argument, as the time spent in the third step on elements in the final result is captured by the $O(k)$ term.

Consider an element $x$ that is a member of one or more input sets, but not part of the result of evaluating the expression. Then there exists some input element $y\ne x$ such that $h(y)=h(x)$.
It is a basic property of universal hash functions that this happens with the same probability as
if $h$ was a truly random function. By our choice of $r$, the probability that this happens for any
particular $x$ is $O(1/w)$. This means that the expected number of such elements is $O(n/w)$,
as desired.

In the case where we are computing an intersection, it suffices to choose $r=\log(\min_i |S_i|)+O(\log w)$, implying the time bound stated in Theorem~\ref{thm:intersect}.

\subsection*{Preprocessing time and space}

The space for the bucketed sets is geometrically increasing with $i$, so it is dominated by the largest
structure which uses $O(n_1)$ words, where $n_1$ is the number of elements in the set. The hash table containing all elements of the set also uses $O(n_1)$ words. The preprocessing time is dominated by
the time for creating the $O(\log w)$ bucketed sets, each in linear time by Lemma~\ref{lem:bucketconvert}. The creation of the hash table takes expected $O(n_1)$ time.

\section{Proofs for section~\ref{sec:succinctsets}}\label{app:lemmas}

{\em Proof of Lemma~\ref{lem:packedoperations}.}
To compute $S_1 \cup S_2$ first merge sorted sequences representing $S_1$ and $S_2$ into a new sequence $X$. Then subtract $X$ by itself shifted $1$ field to the right. The $i$th field in the result stores the value $0$ iff the $i$th field is a duplicate value in $X$. Using this we set all such fields to vacant and compact the resulting packed array thus producing the packed set representation of $S_1 \cup S_2$. By Lemmas~\ref{lem:packedcompact} and \ref{lem:packedmerge} and the fact that the subtraction can be done in $O(1)$ time for each word the result follows. For $S_1 \cap S_2$ we use the same algorithm, with the exception that one of each duplicate fields is set to occupied and all others are set to vacant. \qed

\noindent
{\em Proof of Lemma~\ref{lem:bucketconvert}.}
(1) Consider the case when $b' >b$. First, construct an array of length $2^{b'}$ for the new buckets. Let $B$ be the packed set for a bucket in the representation with parameter $b$. To compute the new representation we need to repartition $B$ into $2^{b' - b}$ new buckets and convert each new bucket into packed sets with parameter $l-b'$. For the repartitioning, observe that $B$ is sorted packed sequence and hence the new buckets are subsequences of $B$. 
It follows that we can traverse and repartition $B$ in $O(2^{b' - b} + |B|\ceil{(l-b)/w})$ time. Converting the packed sets can be done by setting $b' -b$ most significant bits of all fields in the packed sequences to $0$ and then compact the fields as described in the previous section. In total we use $O(2^{b'} + |S|\ceil{(l-b)^2/w})$ time to convert $S$ into a bucketed set with parameter $b'$. The case when $b' < b$ follows similarly, except that packed sets need to be converted to parameter $b'$ representation before they are repartitioned into new buckets. 

(2) Similar to the conversion of the buckets in the proof of (1) we mask out the the $x$ least significant bits of each buckets and compact the fields. \qed

\noindent
{\em Proof of Lemma~\ref{lem:bucketoperations}.}
First, convert $S_1$ and $S_2$ into bucketed sets with parameter $b$ such that $b$ is largest integer satisfying $b\leq \log(|S_1| + |S_2|) - \log w$ using the algorithm from Lemma~\ref{lem:bucketconvert}. Next, perform the desired operation on each of the $2^b$ pairs of buckets using Lemma~\ref{lem:packedoperations} producing a bucketed set $S$. Finally, convert $S$ (if necessary) to a balanced bucketed set.  
\qed 


\section{Improvement of algorithm for asymmetric intersections}

In some situations we can substantially improve the performance of the algorithm described above
by doing a certain transformation of the problem, described in the following. Consider a maximal subexpression that is an intersection of input sets. We denote the number of input sets in the intersection by $m'$. We can improve our algorithm if this intersection is asymmetric in the sense that the smallest of the sets has $n'$ elements, where $n'm'$ is lower than the time needed to compute the (approximate) intersection
in step~1 of the algorithm in Section~\ref{sec:mainalgorithm}. Then we may replace these sets by their intersection by looking up each element in the smallest set in each of the other sets to determine if it
belongs to the intersection. We may create the balanced bucketed set and the hash table needed for the rest of the algorithm in linear time. Thus, the time for this step is $O(n'm')$, potentially reducing the running time.
}

\end{document}